\documentclass[preprint,12pt,sort&compress]{elsarticle}
\usepackage{graphicx}
\usepackage{bm}
\usepackage{citesort}
\usepackage{amsmath}
\usepackage{amssymb}
\journal{Annals of Physics}

\begin{document}
\begin{frontmatter}
\title{Resonant alteration of propagation in guiding structures with complex Robin parameter and its magnetic-field-induced restoration}
\author{O. Olendski}
\ead{oolendski@ksu.edu.sa}
\address{King Abdullah Institute for Nanotechnology, King Saud University, P.O. Box 2454, Riyadh 11451, Saudi Arabia}
\begin{abstract}
Solutions of the scalar Helmholtz wave equation are derived for the analysis of the transport and thermodynamic properties of the two-dimensional disk and three-dimensional infinitely long straight wire in the external uniform longitudinal magnetic field $\bf B$ under the assumption that the Robin boundary condition contains extrapolation length $\Lambda$ with nonzero imaginary part $\Lambda_i$. 
As a result of this complexity, the self-adjointness of the Hamiltonian is lost, its eigenvalues $E$ become complex too and the discrete bound states of the disk characteristic for the real $\Lambda$ turn into the corresponding quasibound states  with their lifetime defined by the eigenenergies imaginary parts $E_i$. Accordingly, the longitudinal flux undergoes an alteration as it flows along the wire with its attenuation/amplification being $E_i$-dependent too. It is shown that, for zero magnetic field, the component $E_i$ as a function of the Robin imaginary part exhibits a pronounced sharp extremum with its magnitude being the largest for the zero real part $\Lambda_r$ of the extrapolation length. Increasing magnitude of $\Lambda_r$ quenches the $E_i-\Lambda_i$ resonance and at very large $\Lambda_r$ the eigenenergies $E$ approach the asymptotic real values independent of $\Lambda_i$. The extremum is also wiped out by the magnetic field  when, for the  large $B$, the energies tend to the Landau levels. Mathematical and physical interpretations of the obtained results are provided; in particular, it is shown that the finite lifetime of the disk quasibound states stems from the $\Lambda_i$-induced currents flowing through the sample boundary. Possible experimental tests of the calculated effect are discussed; namely, it is argued that it can be observed in superconductors by applying to them the external electric field $\cal E$ normal to the surface.
\end{abstract}
\end{frontmatter}

\section{Introduction}
\label{sec1}
A solution $\Psi(\bf r)$ of the scalar Helmholtz wave equation
\begin{equation}\label{WaveEquation1}
{\bm\nabla}^2\Psi({\bf r})+k^2\Psi({\bf r})=0
\end{equation}
is strongly influenced, for the spatially confined domain $\it\Omega$, by the requirements imposed at the sample interface $\it\partial\Omega$. A demand zeroing at the edge a linear superposition of both $\Psi(\bf r)$ and its spatial derivative is known as Robin boundary condition \cite{Gustafson1} with its most general form written as
\begin{equation}\label{Robin1}
\left.{\bf n}{\bm\nabla}\Psi\right|_{\it\partial\Omega}=\left.\frac{1}{\Lambda}\Psi\right|_{\it\partial\Omega},
\end{equation}
$\bf n$ being an inward unit vector normal to the confining interface $\it\partial\Omega$ and Robin parameter $1/\Lambda$ being, in general, a function of the one- (1D), two- (2D) or three-dimensional (3D) radius-vector, $\Lambda\equiv\Lambda(\bf r)$. Such boundary conditions with {\em real} $\Lambda$ emerge naturally in different physical and chemical systems what spurs their theoretical investigation by mathematicians and physicists \cite{Morse1,Strauss1,Balian1,Balian2,Balian3,Fulling1,Ashwin1,Sieber1,Lebedev1,Dancer1,Daners1,Minces1,Solodukhin1,Romeo1,Lebedev2,Berry1,Bordag1,Zayed1,Arendt1,Fulling2,Arendt2,Albuquerque1,Bondurant1,Dowker1,Giorgi1,Baelus1,Pankrashkin1,Liu1,Freitas1,Mintz1,Mintz2,Farina1,Daners2,Giorgi2,Jilek1,Berry2,Giorgi3,Marlettta1,Berry3,Gesztesy1,Teo1,Dean1,Milovanovic1,Olendski5,Behrndt1}. Examples of the vibrating strings and membranes are so well known that they are discussed in the undergraduate textbooks \cite{Strauss1}.  A time-independent Green function method involving a multiple reflection expansion was used for calculating the distribution of eigenvalues of (\ref{WaveEquation1}) for the arbitrary domain with the Robin condition on its surface \cite{Balian1}. Extension to the vector fields and to an arbitrary number of dimensions followed soon \cite{Balian2} (errata of these two papers are fixed in \cite{Balian3}). The requirements of Eq.~(\ref{Robin1}) are used in thermodynamics and statistical physics for the description of the properties of the ideal gases enclosed in the multiply connected bounded containers \cite{Zayed1}. Static and dynamical Casimir effects \cite{Casimir1} were calculated on the straight plates  and miscellaneously curved surfaces \cite{Romeo1,Mintz1,Albuquerque1,Lebedev1,Teo1,Dean1} supporting the Robin condition, Eq.~(\ref{Robin1}). It was also used in the study of the correspondence between a field theory on anti-de Sitter space and a conformal field theory on its boundary \cite{Minces1}. In solid state physics, for the de Broglie particle with mass $m_p$ the wave vector $k$ in (\ref{WaveEquation1}) is expressed as
\begin{equation}\label{WaveVector1}
k=\sqrt{2m_pE}/\hbar,
\end{equation}
where $E$ is a total energy of the particle and $\hbar$ is the reduced Planck constant. The electron in semiconductor nanostructures is satisfactorily well described by the zero Robin distance, $\Lambda=0$, what is a limiting Dirichlet case of the general demand of Eq.~(\ref{Robin1}),
\begin{equation}\label{Dirichlet1}
\left.\Psi\right|_{\it\partial\Omega}=0.
\end{equation}

On the other hand, in the phenomenological Ginzburg-Landau (GL) theory of superconductivity \cite{Ginzburg1,deGennes1} the Cooper pair behavior is described by the order parameter $\Psi(\bf r)$ with its square being equal to the density of the superconducting particles $n_s$, 
\begin{equation}\label{density1}
n_s=|\Psi({\bf r})|^2.
\end{equation}
In general, the order parameter satisfies nonlinear GL equation
\begin{equation}\label{NonlinearEquation1}
{\bm\nabla}^2\Psi({\bf r})+k^2\Psi({\bf r})+\beta\left|\Psi({\bf r})\right|^2\Psi({\bf r})=0.
\end{equation}
Here, the role of the energy $E$ in (\ref{WaveVector1}) is played by the parameter $-\alpha$:
\begin{equation}\label{Energy1}
E\equiv -\alpha=\frac{\hbar^2}{2m_p\xi^2\left(0\right)}\left(1-\frac{T}{T_c}\right),
\end{equation}
with $T$ being the actual temperature of the superconducting material, $T_c$ being the bulk critical temperature at zero magnetic field, $\xi\left(0\right)$ representing zero-temperature coherence length, and Cooper pair mass being equal to the double bare electron mass $m_e$. An expression for the second GL parameter $\beta$ in Eq.~(\ref{NonlinearEquation1}) is derived from the microscopic equations of superconductivity, and for the "pure" superconductors it reads
\begin{equation}\label{beta1}
\beta\simeq\frac{1}{N\left(0\right)}\frac{\hbar^2}{2m_p\xi^4\left(0\right)}\frac{1}{\left(k_BT_c\right)^2}
\end{equation}
with $N\left(0\right)$ being the density of states at the Fermi energy and $k_B$ the Planck constant. For the 'dirty' superconductors (alloys) $\xi^2(0)$ in GL parameters $\alpha$ and $\beta$ should be substituted by $\xi\left(0\right)l$, with $l$ being the mean free path of the material \cite{deGennes1}. In the temperature range close to the transition to the normal state, the order parameter is small, and the cubic term can be safely neglected \cite{deGennes1} thus simplifying the nonlinear GL equation (\ref{NonlinearEquation1}) into its linear Helmholtz counterpart, Eq.~(\ref{WaveEquation1}), which is much easier to handle. As it follows from (\ref{beta1}), the influence of the cubic term can be also strongly suppressed by the appropriate change of the materials \cite{Olendski5,Olendski6} which leads again to the analysis of (\ref{WaveEquation1}) whose smallest eigenvalue $E_{min}$, according to (\ref{Energy1}), defines the the sample temperature \cite{deGennes1}:
\begin{equation}\label{Temperature1}
T=T_c\left[1-\frac{2m_p\xi^2\left(0\right)}{\hbar^2}E_{min}\right].
\end{equation}
This expression shows an enormous importance of minimizing the lowest eigenenergy $E_{min}$ since this leads to the higher superconductor temperatures. One of the possible methods of this minimization is the curving of the previously straight sample \cite{Olendski4,Montevecchi1,Montevecchi2,Olendski6}.

In the GL theory the order parameter behavior near the surface is governed by the same Eq.~(\ref{Robin1}).  The de Gennes distance, as is the length $\Lambda$ called in this field of physical research, is a quantitative measure of the interaction between the superconductor and the ambient environment. For example, the border with the vacuum or the insulator is described by the  infinite extrapolation length, $1/\Lambda=0$, which leads to the Neumann limit of Eq.~(\ref{Robin1}),
\begin{equation}\label{Neumann1}
\left.\frac{\partial\Psi}{\partial n}\right|_{\it\partial\Omega}=0.
\end{equation}
For the superconductor border with the normal metal the de Gennes distance takes  the values in the interval $0<\Lambda<\infty$ where the limit of Eq.~(\ref{Dirichlet1}) is approached for the bordering ferromagnets. Solutions of (\ref{NonlinearEquation1}) were found and analysed for the different domains $\Omega$ and Robin factors $\Lambda$ \cite{Zaitsev2,Andryushin1,Andryushin2,Lu1}; in particular, a crucial role of the extrapolation length in the nucleation of the superconductivity in thin films was underlined. Interestingly, the de Gennes distance can take negative values, $\Lambda<0$, for the boundary with the superconductor with higher critical temperature $T_c$. This theoretically predicted enhancement \cite{Fink1,Montevecchi1,Montevecchi2} was indeed observed in cold worked In$_{0.993}$Bi$_{0.007}$ foils \cite{Fink1} and tin samples \cite{Kozhevnikov1}. It was estimated that in the former case the extrapolation length is $\Lambda\sim -1$ $\mu$m \cite{Montevecchi1}. Negative extrapolation length appears also in the theory of twinning plane superconductivity \cite{Khlyustikov1}. Robin boundary conditions, Eq.~(\ref{Robin1}), are also utilized in the study of magnetic multilayers \cite{Olendski5} and ferrite-filled waveguides \cite{Gurevich1,Bosma1,Bosma2,Fay1,BadenFuller1}. Recently, they were also used to model the tunneling current and the optical gain of GaAs/Al$_x$Ga$_{1-x}$As quantum cascade lasers \cite{Milovanovic1}.

To take into account the influence on the charged carriers of the external magnetic field $\bf B$, one needs in a standard way to incorporate it into the Helmholtz equation through the introduction of the vector potential $\bf A$,
\begin{equation}\label{VectorPotential1}
{\bf B}={\bm\nabla}\times{\bf A}.
\end{equation}
Then, for the de Broglie particle, Eq.~(\ref{WaveEquation1}) turns to
\begin{equation}\label{Magnetic1}
\left({\bm\nabla}-i\frac{q}{\hbar}{\bf A}\right)^2\Psi({\bf r})+k^2\Psi({\bf r})=0,
\end{equation}
where $q$ is the particle charge; in particular, for the Cooper pair, $q=-2e$, with $e$ being the absolute value of the electronic charge. Eq.~(\ref{NonlinearEquation1}) is modified in the same way:
\begin{equation}\label{NonlinearEquation2}
\left({\bm\nabla}-i\frac{q}{\hbar}{\bf A}\right)^2\Psi({\bf r})+k^2\Psi({\bf r})+\beta\left|\Psi({\bf r})\right|^2\Psi({\bf r})=0.
\end{equation}
In the GL theory, the boundary condition changes to
\begin{equation}\label{Robin2}
\left.{\bf n}\left({\bm\nabla}-i\frac{q}{\hbar}{\bf A}\right)\Psi\right|_{\it\partial\Omega}=\left.\frac{1}{\Lambda}\Psi\right|_{\it\partial\Omega}.
\end{equation}
Theoretical analysis of the combined influence of the magnetic field and real Robin length on the superconductors \cite{Hurault1,Takacs1,Takacs2,Takacs3,Takacs4,Buisson1,Masale1,Richardson1,Hornberger1,Baelus2,Pogosov1,Hornberger2,Pan1,Slachmuylders1,Kachmar1,Zhu1,Kachmar2,Kachmar3,Kachmar4,Kachmar5,Zha1,Barba1,Barba-Ortega1,Barba-Ortega2} revealed, among other features, a significant influence of the parameter $\Lambda$ on the nucleation of the superconductivity, critical magnetic fields and localization properties of the superconducting state.

A natural extension of the {\it real} Robin distance is its continuation into the whole {\it complex} plane $\Lambda\equiv(\Lambda_r,\Lambda_i)$. For example, the boundary conditions with purely imaginary $\Lambda=i\Lambda_i$ for the 1D and 2D geometries are extensively used \cite{Krejcirik1,Krejcirik2,Krejcirik3,Borisov1,Krejcirik4} in the actively developing field of $\cal PT$-symmetric quantum mechanics \cite{Bender1,Bender2,Bender3}. Apart from the purely theoretical interest \cite{Davies1}, the analysis of the systems with the complex Robin lengths has large practical applications as they model scattering phenomena in different media \cite{Lakshtanov1}. Consider, for example, electromagnetic or acoustical processes when the wave vector $k$ from (\ref{WaveEquation1}) is the ratio of the actual frequency $\omega$ and the speed $c$ of the propagation of the corresponding oscillations:
\begin{equation}\label{WaveVector2}
k=\frac{\omega}{c}.
\end{equation}
\begin{figure}
\centering
\includegraphics[width=0.99\columnwidth]{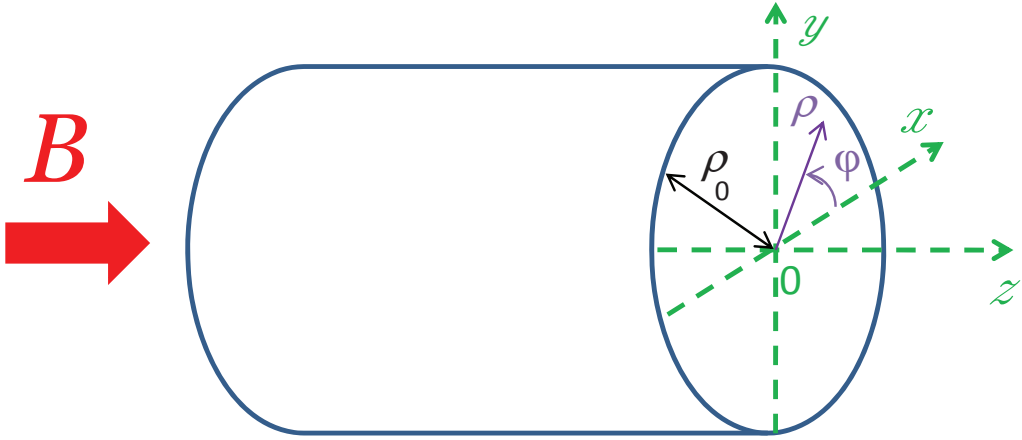}
\caption{\label{Fig1}
Schematic of the infinitely long straight  superconductor wire of the radius $\rho_0$ subjected to the uniform magnetic field $\bf B$ parallel to its axis. Channel walls support Robin boundary condition, Eq.~(\ref{Robin1}), with complex extrapolation length $\Lambda$ uniform along the wire surface. Cartesian $(x,y,z)$ and cylindrical $(\rho,\varphi,z)$ systems of coordinates are also shown with their origins lying on the waveguide axis. Curved arrow shows the azimuthal direction in which the polar angle $\varphi$ grows.
}
\end{figure}
It is well known that a porous lining of the walls of the sound duct results in nonzero and, in general, complex acoustical admittance $Y$ \cite{Allard1}. Corresponding boundary condition for the air pressure $p$ that satisfies  Eq.~(\ref{WaveEquation1}) is then described by (\ref{Robin1}) with the coefficient $1/\Lambda$ being proportional to the admittance $Y$ \cite{Grigoryan1,Grigoryan2,Ko1,Ko2,Rostafinski1,Rienstra1,Rienstra2,Felix1,Bi1}. The same boundary requirements \cite{Weston1} appear in the study of the propagation and scattering processes in the impedance electromagnetic waveguides with the fields independent of the tangential direction when the division onto the $TM$- and $TE$-modes is preserved \cite{Katsenelenbaum1}. In this case, the coefficient $1/\Lambda$ in (\ref{Robin1}) is either proportional or inverse proportional to the complex surface impedance $w$ and its imaginary part is determined by the real component of the impedance \cite{Katsenelenbaum1}. Apparently, the same model with complex extrapolation length $\Lambda$ describes the processes in the ferrite-filled guiding structures with complex permeability \cite{Krupka1,Bosma1,Bosma2,Fay1}. Dissipative Schr\"{o}dinger operators with complex 1D Robin boundary conditions were analyzed for the study of the drift-diffusion phenomena in low-dimensional semiconductors \cite{Kaiser1,Kaiser2,Kaiser3,Baro1}. Superconducting de Gennes distance can take complex values too. Namely, it was argued recently \cite{Lipavsky1,Morawetz1} that the electric field $\mbox{\boldmath$\cal E$}$ applied perpendicularly to the surface should be accounted for in the total de Gennes distance $\Lambda_{tot}$ by the addition to the inverse zero-field extrapolation length $1/\Lambda$ of the extra term ${\cal E}/U_s$
\begin{equation}\label{TotalExtrapolationLength1}
\frac{1}{\Lambda_{tot}}=\left.\frac{1}{\Lambda}\right|_{{\cal E}=0}+\frac{\cal E}{U_s}
\end{equation}
with the potential $U_s$ being expressed through the parameters of the GL theory:
\begin{equation}\label{PotentialU}
\frac{1}{U_s}=\eta\kappa^2\frac{\partial\ln T_c}{\partial\ln n_s}\frac{\left|q\right|\epsilon_s}{m_pc^2}.
\end{equation}
Here, the reduction factor $\eta$ is the ratio of the band gap extrapolated to the surface to the gap value at the surface; dimensionless GL parameter $\kappa$ is the ratio of the zero-temperature London penetration length $\lambda(0)$ to the coherence length, $\kappa=\lambda(0)/\xi(0)$; $\epsilon_s$ is superconductor ionic background permittivity, and $c$ is speed of light.  According to the estimates \cite{deGennes1}, the coefficient $\eta$ is of the order of unity. Electric field influence on superconductors is still an open question, see Ref. \cite{Kolacek1} and references therein. If the permittivity $\epsilon_s$ has a noticeable imaginary part, so does, according to (\ref{TotalExtrapolationLength1}) and (\ref{PotentialU}), the total extrapolation length too. It is well known that the real value of the de Gennes distance is due to the fact that no superconducting current flows through its surface \cite{deGennes1}; indeed, as we shall show below, a complex part of the extrapolation length forces the current to acquire nonzero normal (with respect to the interface) component.

\begin{figure}
\centering
\includegraphics[width=0.75\columnwidth]{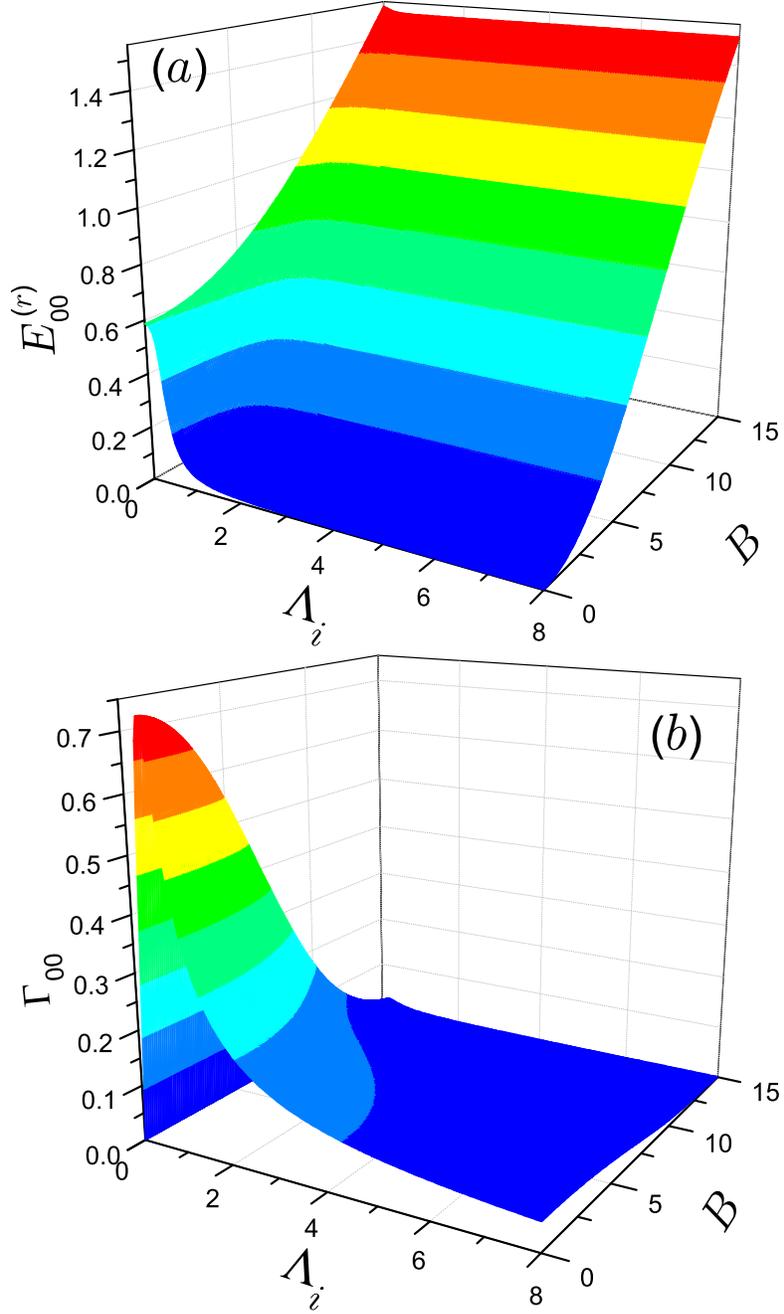}
\caption{\label{Fig2}
(a) Real $E_{00}^{(r)}$ and (b) negative double imaginary $\Gamma_{00}$ parts of the total transverse energy $E_{00}^\perp$ as functions of $\Lambda_i$ and $B$ for $\Lambda_r=0$.
}
\end{figure}

Thus, solutions of the Helmholtz equation with {\it complex} Robin parameter are applicable in many areas of physics. Even though some investigations have tackled different aspects of this problem, as the above review demonstrates, so far no comprehensive detailed dependence of the bounded domain properties on the extrapolation length $\Lambda$ was given and no magnetic field influence was addressed. Present research closes this gap for the simplest case of the 3D straight channel and its cross-sectional 2D disk; namely, the problem of the infinitely long straight waveguide of the circular shape with the boundary conditions, Eq.~(\ref{Robin1}), supporting complex extrapolation length $\Lambda$ is formulated and solved. Uniform magnetic field {\bf B} is pointed in the wire axis direction. As stated above, without magnetic field such a model and its analysis based on the solutions of Eq.~(\ref{WaveEquation1}) is relevant for the description of the wave dynamics in lined acoustical and impedance electromagnetic ducts as well as ferrite or superconductor wires. Turning on the field confines the applicability of the model to the latter situation when Eq.~(\ref{Magnetic1}) comes under scrutiny. In either case, analytical solutions of the corresponding equation are derived and analyzed. Based on these solutions, currents and magnetizations are calculated in the whole complex de Gennes plane $\Lambda_r-\Lambda_i$. It is shown that the magnetization for the purely imaginary extrapolation lengths exhibits  basically the same dependence as for the real Robin distances while the longitudinal current along the wire shows resonance features in its dependence on the imaginary $\Lambda$. The physical and mathematical reasons for such highly nonmonotonic behavior are described in terms of the transverse currents induced by the imaginary component of the extrapolation length. These currents flow through the superconductor boundary and in this way they change the number of the Cooper pairs participating in the longitudinal transport. Possible experimental confirmations of the developed theory are discussed.

The paper is organized as follows. In Section II our model is presented and a necessary formulation of our method is given. Section III is devoted to the presentation and detailed mathematical and physical interpretation of the calculated results. Summary of the research with possible future extensions is provided in Section IV.

\begin{figure}
\centering
\includegraphics[width=0.75\columnwidth]{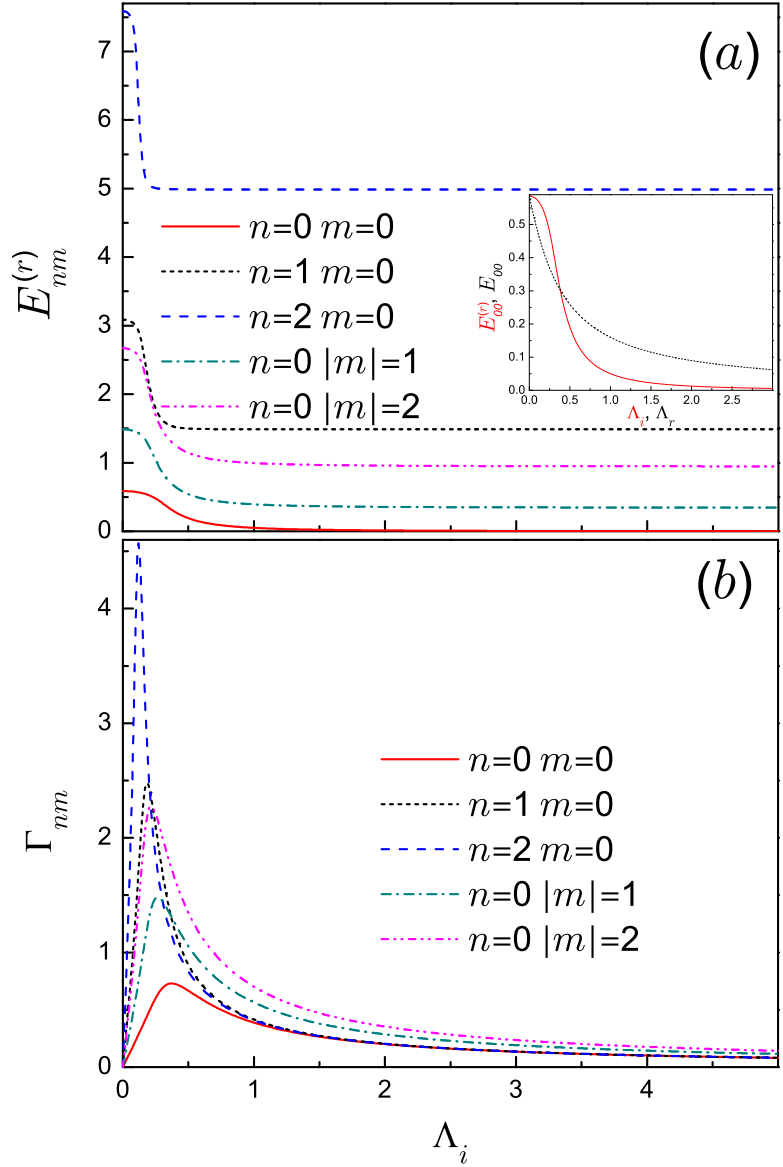}
\caption{\label{Fig3}
(a) Real $E_{nm}^{(r)}$ and (b) negative double imaginary $\Gamma_{nm}$ parts of the transverse energy $E_{nm}^\perp$ as functions of $\Lambda_i$ for $\Lambda_r=0$ and zero magnetic field for several radial $n$ and azimuthal $m$ quantum numbers. Solid lines depict the state with $n=m=0$, dotted lines are for $n=1$ and $m=0$, dashed lines - for $n=2$ and $m=0$, dash-dotted lines - for $n=0$ and $|m|=1$, and dash-dot-dotted lines - for $n=0$ and $|m|=2$. The inset in panel (a) shows real energy $E_{00}$ as a function of the real extrapolation length $\Lambda_r$ (dotted curve) superimposed on the dependence of the real part $E_{00}^{(r)}$ of the complex energy $E_{00}^\perp$ versus purely imaginary extrapolation length $i\Lambda_i$ (solid line).
}
\end{figure}

\section{Model and formulation}
\label{sec2}

Infinitely long straight wire of the circular cross section with the radius $\rho_0$ is placed into the uniform magnetic field $\bf B$ with its direction coinciding with the channel axis (Fig. \ref{Fig1}). Cylindrical walls of the waveguide support boundary condition, as described by Eq.~(\ref{Robin1}) with the uniform along the length and circumference extrapolation length $\Lambda$. We do not confine the value of $\Lambda$ to be real concentrating on the properties of the structure at the complex de Gennes distances. Geometry of the system dictates a natural choice of the cylindrical system of coordinates $(\rho,\varphi,z)$ with its origin lying at the circle center and the $z$ axis being parallel  to the waveguide. We will seek the solutions of Eq.~(\ref{Magnetic1}) with the vector potential from Eq.~(\ref{VectorPotential1}) written in the symmetrical gauge,
\begin{equation}\label{VectorPotential2}
{\bf A}=(0,B\rho/2,0).
\end{equation} 
The fact that we treat Eq.~(\ref{Magnetic1}) with the uniform field means that, for the case of superconductors, we restrict our consideration to the range of the magnetic intensities and temperatures close to the transition between superconducting and normal states \cite{deGennes1} even though the results obtained are covered by the full GL theory (see below) and have much wider validity range. We will operate with the energy $E$ through which the wave vector $k$ is expressed, according to (\ref{WaveVector1}). Such a treatment describes a superconductor wire. A transition to the frequency $\omega$ of the acoustical or electromagnetic oscillations can be readily done with the help of (\ref{WaveVector2}). It is important to remark here that, for the case of ferrites, the static magnetic field influence on the propagation properties is through the dependence of the magnetic permeability tensor $\mu$ on $B$ and, thus, one needs to consider Eq.~(\ref{WaveEquation1}) where the speed of propagation depends on the magnetic field through $\mu$ \cite{Gurevich1,BadenFuller1}. In order to make the obtained results  as generic as possible, it is convenient to express all quantities in dimensionless units. In the GL theory, one usually chooses as a unit of distance the coherence length $\xi(0)$; however, to make our results directly applicable in the acoustics and electrodynamics, we choose the radius $\rho_0$ as a natural unit of length; accordingly, if not stated otherwise, all energies will be measured in units of the ground-state energy $\pi^2\hbar^2/(2m_p\rho_0^2)$ of the infinite Dirichlet 1D quantum well of width $\rho_0$; all momenta, in units $1/\rho_0$; magnetic fields, in units of $\hbar/(|q|\rho_0^2)$; magnetization, in units of $\hbar|q|/(2m_p)$; 2D current density, in units of $q\hbar/(m_p\rho_0^4)$; current, in units of $q\hbar/(m_p\rho_0)$; time, in units of $2m_p\rho_0^2/(\pi^2\hbar)$. In these units Eq.~(\ref{WaveVector1}) transforms to $k=\pi\sqrt{E}$. Then, Eq.~(\ref{Magnetic1}) for the order parameter $\Psi(\rho,\varphi,z)$ becomes
\begin{equation}\label{WaveEquation2}
\frac{1}{\rho}\frac{\partial}{\partial\rho}\left(\rho\frac{\partial\Psi}{\partial\rho}\right)+\frac{1}{\rho^2}\frac{\partial^2\Psi}{\partial\varphi^2}+iB\frac{\partial\Psi}{\partial\varphi}-\frac{1}{4}B^2\rho^2\Psi+\frac{\partial^2\Psi}{\partial z^2}+\pi^2E\Psi=0
\end{equation}
with $E$ being a total energy of the particle. Factoring out the $z$-dependence
\begin{equation}\label{Factoring1}
\Psi(\rho,\varphi,z)=e^{ik_zz}\psi(\rho,\varphi)
\end{equation}
leads to the equation for the transverse function $\psi(\rho,\varphi)$:
\begin{equation}\label{WaveEquation3}
\frac{1}{\rho}\frac{\partial}{\partial\rho}\left(\rho\frac{\partial\psi}{\partial\rho}\right)+\frac{1}{\rho^2}\frac{\partial^2\psi}{\partial\varphi^2}+iB\frac{\partial\psi}{\partial\varphi}-\frac{1}{4}B^2\rho^2\psi+\pi^2E^\perp\psi=0.
\end{equation}
Longitudinal wave vector $k_z$ and the transverse energy $E^\perp$ are related as
\begin{equation}\label{relation1}
k_z=\pi\sqrt{E-E^\perp}.
\end{equation}

Rotational symmetry of the system allows one to separate out the angular and radial dependencies in the transverse function $\psi(\rho,\varphi)$:
\begin{equation}\label{Factoring2}
\psi_{nm}(\rho,\varphi)=\frac{1}{\sqrt{2\pi}}e^{im\varphi}R_{nm}(\rho).
\end{equation}
Here, $m=0,\pm 1,\pm 2,\ldots$ and $n=0,1,\ldots$ are the azimuthal and the principal quantum numbers, respectively. In this way one arrives at the equation for the radial function $R_{nm}(\rho)$:
\begin{equation}\label{RadialEquation1}
\frac{d^2R_{nm}}{d\rho^2}+\frac{1}{\rho}\frac{dR_{nm}}{d\rho}-\left(\frac{m}{\rho}+\frac{1}{2}B\rho\right)^2R_{nm}+\pi^2E_{nm}^\perp R_{nm}=0.
\end{equation}
This equation is supplemented by the boundary condition for the function $R_{nm}(\rho)$. Our choice of the vector potential in the form of (\ref{VectorPotential2}) drops it out from (\ref{Robin2}) which becomes
\begin{equation}\label{Robin3}
\left.\left(\frac{dR_{nm}}{d\rho}+\frac{1}{\Lambda}R_{nm}\right)\right|_{\rho=1}=0.
\end{equation}
Eqs. (\ref{RadialEquation1}) and (\ref{Robin3}) constitute the problem of finding the eigenfunctions $R_{nm}(\rho)$ and eigenenergies $E^\perp_{nm}$ of the 2D circular disk with its circumference supporting the boundary condition with, in general, complex $\Lambda$. We mention that the field-free case of the real constant \cite{Sieber1,Berry2} or varying \cite{Berry2,Marlettta1,Berry3} along the perimeter extrapolation length was considered before, as was the case of the quantum dot with real constant $\Lambda$ in the uniform magnetic field \cite{Takacs3,Takacs4,Buisson1} with the  Dirichlet \cite{Dingle1,Dingle2,Nakamura1,Geerinckx1,Constantinou2,Avishai1} or Neumann \cite{SaintJames1,Dalmasso1,Constantinou1,Buisson1,Moshchalkov2,Benoist1} limits.

Analytical solution to Eq.~(\ref{RadialEquation1}) is expressed via the Kummer confluent hypergeometric function $M(a,b;x)$ \cite{Abramowitz1}:
\begin{eqnarray}
&&R_{nm}(\rho)=\gamma_{nm}\exp\left(-\frac{1}{4}B\rho^2\right)\left(\frac{1}{2}B\rho^2\right)^{|m|/2}\nonumber\\
\label{RadialFunction1}
&&\times M\left(\frac{m+|m|+1}{2}-\frac{\pi^2}{2}\frac{E_{nm}^\perp}{B},|m|+1;\frac{1}{2}B\rho^2\right)
\end{eqnarray}
with the real coefficient $\gamma_{nm}$ determined from the normalization condition which is either of the form from (\ref{density1}) for superconductors or  
\begin{equation}\label{normalization1}
\int_0^1\left|R_{nm}(\rho)\right|^2\rho d\rho=1
\end{equation}
for the electromagnetic and acoustical waveguides. Applying boundary requirement, Eq.~(\ref{Robin3}), one arrives at the following transcendental equation for the determination of the energies $E^\perp_{nm}$:
\begin{eqnarray}
&&\left(\left|m\right|-\frac{B}{2}+\frac{1}{\Lambda}\right)M\left(\frac{m+|m|+1}{2}-\frac{\pi^2}{2}\frac{E_{nm}^\perp}{B},\left|m\right|+1;\frac{B}{2}\right)\nonumber\\
\label{MagneticEq1}
&&+BM'\left(\frac{m+|m|+1}{2}-\frac{\pi^2}{2}\frac{E_{nm}^\perp}{B},\left|m\right|+1;\frac{B}{2}\right)=0
\end{eqnarray}
with prime denoting a derivative of the function with respect to the last argument.  Instead of the derivative, one can use the Kummer function itself with the different parameters according to \cite{Abramowitz1}:
$$
M'(a,b,z)=\frac{a}{b}M(a+1,b+1,z).
$$
Utilizing properties of the Kummer function \cite{Abramowitz1}, it is elementary to show that in the limit of strong magnetic field,  $B\rightarrow\infty$, solutions of (\ref{MagneticEq1}) tend from below to
\begin{equation}\label{LandauLevels1}
E_{nm}^\perp=\frac{2}{\pi^2}\left(n+\frac{m+\left|m\right|+1}{2}\right)B
\end{equation}
which are familiar Landau levels \cite{Page1,Landau2}. In the opposite limit of the vanishing magnetic field, $B\rightarrow 0$, Eq.~(\ref{MagneticEq1}) transforms to
\begin{equation}\label{BesselEq1}
\pi\sqrt{E_{nm}^\perp}J_{|m|}'\left(\pi\sqrt{E_{nm}^\perp}\right)+\frac{1}{\Lambda}J_{|m|}\left(\pi\sqrt{E_{nm}^\perp}\right)=0
\end{equation}
with $J_{m}(x)$ being a Bessel function of the order $m$ \cite{Abramowitz1}. This equation was derived before for the case of the real \cite{Sieber1,Romeo1,Berry2,Olendski5,Bosma1,Bosma2} and complex \cite{Rienstra1} extrapolation lengths. Zero-field wave function is expressed then as
\begin{equation}\label{RadialFunction2}
R_{nm}(\rho)=\gamma_{nm}J_{|m|}\left(\pi\sqrt{E_{nm}^\perp}\rho\right).
\end{equation}
Of course, this last expression could be obtained directly from (\ref{RadialEquation1}) if one puts $B=0$ in it and, subsequently, Eq.~(\ref{BesselEq1}) is derived from it by means of (\ref{Robin3}). From (\ref{MagneticEq1}) and (\ref{BesselEq1}) it straightforwardly follows that for the complex extrapolation lengths, $\Lambda=\Lambda_r+i\Lambda_i$ with real $\Lambda_r={\rm Re}(\Lambda)$ and $\Lambda_i={\rm Im}(\Lambda)$, the energies are complex too,
\begin{equation}\label{ComplexEnergy1}
E_{nm}^\perp=E_{nm}^{(r)}-i\frac{\Gamma_{nm}}{2}
\end{equation}
with real $E_{nm}^{(r)}$ and $\Gamma_{nm}$, and the following property holds:
\begin{equation}\label{conjugate1}
E_{nm}^\perp\left(\overline{\Lambda}\right)=\overline{E_{nm}^\perp\left(\Lambda\right)}
\end{equation}
with the overline denoting a complex conjugate value. The same is true for the corresponding radial functions too. Minus sign and the fraction in the imaginary part of the right-hand side of (\ref{ComplexEnergy1}) were taken to be consistent with the scattering theory \cite{Landau2,Newton1} which teaches that an imposing of the {\it complex} boundary conditions eliminates the self-adjointness of the corresponding Hamiltonian what results in the {\it complexity} of its eigenenergies. Depending on the type of the requirement at the border (in our particular case, the sign of $\Lambda_i$), the imaginary part can be either positive, $\Gamma<0$, or negative, $\Gamma>0$. In the latter case, the true bound level with the discrete at $\Lambda_i=0$ energy transforms into the quasibound state broadened over the width $\Gamma$ around its central value $E^{(r)}$. The particle does not stay forever in such quasibound state but after the time 
\begin{equation}\label{lifetime1}
\tau_{nm}=\frac{1}{\Gamma_{nm}}
\end{equation}
leaves the system.  As (\ref{lifetime1}) shows, the lifetime $\tau_{nm}$ is inversely proportional to the half width. The negative value of $\Gamma$ means, in turn, a flux of the particles into the system and their accommodation inside it with the temporal accommodation rate being proportional to $1/|\Gamma|$. We will discuss these two situations in more detail later in the text.

It is important to state that even though (\ref{MagneticEq1}) and (\ref{BesselEq1}) were derived for the linearized GL equation, it can be shown that for the thin superconducting disks they follow also from the complete nonlinear GL theory. For doing this, one needs to employ the method of minimizing of the free energy $F$ of the superconducting state with respect to the coefficient linking the order parameter of the nonlinear equation with its linear counterpart, Eq.~(\ref{Factoring2}). As this procedure is described in detail for the superconducting rings \cite{Zhu1,Zha1}, we do not repeat it here referring the reader to the original research, Refs. \cite{Zhu1,Zha1}.

Knowledge of the eigenfunctions and eigenenergies allows one to find other physical characteristics of the structure. For example, in the case of superconductors a magnetization operator
\begin{equation}\label{Magnetization1}
\hat{M}=i\frac{\partial}{\partial\varphi}-\frac{1}{2}B\rho^2
\end{equation}
is used to calculate the magnetic moment $M_z$ of the 2D disk:
\begin{equation}\label{Magnetization2}
M_z=\langle\psi_{nm}(\rho,\varphi)|\hat{M}|\psi_{nm}(\rho,\varphi)\rangle.
\end{equation}
One immediately gets:
\begin{equation}\label{Magnetization3}
M_z=-\left(m+\frac{B}{2}\int_0^1\left|R_{nm}(\rho)\right|^2\rho^3 d\rho\right).
\end{equation}
It is worthwhile to note that the zero-field case allows analytical calculation of  the integrals in (\ref{normalization1}) and (\ref{Magnetization3}) \cite{Gradshteyn1,Prudnikov1} for the real extrapolation lengths when the radial functions are real too; however, since there are no in known to us literature \cite{Abramowitz1,Gradshteyn1,Prudnikov1,Prudnikov2,Brychkov1} analytical expressions for these integrals with the complex functions $R_{nm}(\rho)$, their direct numerical quadrature was performed. The same procedure was used for the integrals involving the Kummer functions for the nonzero magnetic fields with either real or complex extrapolation length.

An expression for the superconductor current density $\bf j$ is written as \cite{Ginzburg1,deGennes1}:
\begin{equation}\label{current1}
{\bf j}={\rm Im}\left[\overline{\Psi}\left(\bf r\right){\bm\nabla}\Psi\left(\bf r\right)\right]+{\bf A}\overline{\Psi}\left(\bf r\right)\Psi\left(\bf r\right).
\end{equation}
Similar formula (with, of course, ${\bf A}=0$) can be used for the Poynting vector in electrodynamics \cite{Jackson1} or for the sound energy density flux in acoustics \cite{Landau1}. In our configuration, for the 2D quantum dot without the $z$ dependence it transforms to
\begin{equation}\label{current2}
{\bf j}=\frac{1}{2\pi}\left[{\rm Im}\left(\overline{R}\frac{dR}{d\rho}\right){\bf e}_\rho+\left(\frac{m}{\rho}+\frac{1}{2}B\rho\right)\left|R\right|^2{\bf e}_\varphi\right]
\end{equation}
with the unit orthogonal vectors ${\bf e}_\rho$ and ${\bf e}_\varphi$ along the radial and azimuthal directions, respectively. The real value of the extrapolation length $\Lambda$ guarantees that no supercurrent flows through the border \cite{deGennes1}. This general statement can be checked directly with the help of (\ref{current1}) and (\ref{Robin2}) which in our dimensionless units reads
\begin{equation}\label{Robin4}
\left.{\bf n}\left({\bm\nabla}+i{\bf A}\right)\Psi\right|_{\it\partial\Omega}=\left.\frac{1}{\Lambda}\Psi\right|_{\it\partial\Omega}.
\end{equation}
For our system, Eq.~(\ref{current2}) vividly proves it since the normal component of the current described by the term at the radial unit vector, for the real radial functions is an identical zero at the cylindrical surface, what is the case for the real Robin coefficient, as Eq.~(\ref{Robin3}) demonstrates.

For the wire the total longitudinal supercurrent $J_z$ is given as
\begin{equation}\label{TotalZCurrent}
J_z=\int_0^1\rho d\rho\int_0^{2\pi}d\varphi j_z={\rm Re}(k_z)\exp\left[-2{\rm Im}(k_z)z\right].
\end{equation}
Eq.~(\ref{density1}) shows that the longitudinal dependence of the density $n_s$ is of the same form:
\begin{equation}\label{density2}
n_s=\left|R\left(\rho\right)\right|^2\exp\left[-2{\rm Im}(k_z)z\right].
\end{equation}
Expressions for ${\rm Re}(k_z)$ and ${\rm Im}(k_z)$ follow straightforwardly from (\ref{relation1}) and (\ref{ComplexEnergy1}):
\begin{subequations}\label{Wavevector3}
\begin{eqnarray}\label{Wavevector3Real}
{\rm Re}(k_z)&=&\frac{\pi}{\sqrt{2}}\sqrt{\sqrt{\left[E-E_{nm}^{\left(r\right)}\right]^2+\left(\Gamma_{nm}/2\right)^2}+\left[E-E_{nm}^{\left(r\right)}\right]}\\
\label{Wavevector3Imag}
{\rm Im}(k_z)&=&\frac{\pi}{\sqrt{2}}\frac{\Gamma_{nm}/2}{\sqrt{\sqrt{\left[E-E_{nm}^{\left(r\right)}\right]^2+\left(\Gamma_{nm}/2\right)^2}+\left[E-E_{nm}^{\left(r\right)}\right]}}.
\end{eqnarray}
\end{subequations}
Note that the real and imaginary parts of the wave vector $k_z$ are not independent but are related to each other as
\begin{equation}
{\rm Re}(k_z){\rm Im}(k_z)=\frac{\pi^2}{4}\Gamma.
\end{equation}

As is seen from (\ref{TotalZCurrent}) and~(\ref{Wavevector3}), for the real extrapolation lengths when, as stated above, $\Gamma_{nm}=0$, the supercurrent, or the electromagnetic or acoustic energy flows along the channel without changes in its intensity, ${\rm Re}(k_z)=\pi\sqrt{E-E_{nm}^{\left(r\right)}}$, ${\rm Im}(k_z)=0$; however, as soon as the Robin distance acquires a complex component, the nonzero imaginary part of the wave vector forces the current to change as it flows down the waveguide. For the superconducting, electromagnetic or sound transport phenomena a positive sign of $\Gamma$ determines an attenuation of the corresponding energy flux and the particle density with the fading intensity being proportional to its magnitude, as Eqs.~(\ref{TotalZCurrent})-(\ref{Wavevector3}) demonstrate. However, in our case, as Eq.~(\ref{conjugate1}) manifests, the value of $\Gamma$ can be negative too since the sign of the imaginary part of the total extrapolation length $\Lambda_{tot}$ in our model can be changed by the simple switching of the gate voltage, as it follows from (\ref{TotalExtrapolationLength1}) and (\ref{PotentialU}). In this case, instead of dissipating the energy into the surrounding environment, as is the case for $\Gamma>0$, the channel absorbs it from outside. In reality, the increase of the current approaches some saturation value since an infinite pumping of the energy is physically senseless. This saturation is not covered by our model. Moreover, in the superconductor wire the current upon reaching some critical value destroys the superconductivity. This, too, should be covered by more elaborated theories.

Used in the present research GL theory, being phenomenological in its nature, does not provide an explanation of the microscopic mechanism of the carrier transformation in the superconducting and nonsuperconducting media. Invoking the Bardeen-Cooper-Schrieffer theory \cite{deGennes1} and our earlier discussion about the electric field influence on the Robin parameter [see (\ref{TotalExtrapolationLength1}) and (\ref{PotentialU})], one can say that single electrons pushed by the appropriately directed perpendicular electric field from, say, normal metal into the superconductor, are bound inside it into the Cooper pair and, thus, make an additional contribution to the total longitudinal supercurrent as the nonsuperconducting flow of the normal electrons is an identical zero in the thermal equilibrium. Switching the direction of the electric field forces the Cooper pair to move out of the superconductor into the medium that can not support this bound state with the corresponding decrease of the longitudinal supercurrent. In the same way, the electrodynamics operating with the complex impedance $w$ \cite{Weston1,Katsenelenbaum1} does not provide the detailed microscopic description of the wave attenuation which is explained  by the electromagnetic energy transformation into the joule motion in the waveguide metallic walls. These nuances are simply incorporated into the value of $w$.

To get more insight into the physical meaning of the complex energy, it is instructive to calculate the 2D divergence of the supercurrent in the quantum dot. Eq.~(\ref{current2}) shows that only the radial component contributes. Utilizing properties of the Bessel functions, one gets:
\begin{equation}\label{divergence1}
{\rm div}{\bf j}=\frac{\pi}{4}|R|^2\Gamma.
\end{equation}
For the waveguide this expression in each cross section $z={\rm const}$ should be multiplied by the familiar from above exponential factor $\exp\left[-2{\rm Im}(k_z)z\right]$. Thus, the imaginary part of the energy defines the divergence of the current with its positive (negative) value meaning that the corresponding spatial point serves as a sink (source). This explains the longitudinal current change for the complex energies as for the positive (negative) $\Gamma$ the Cooper pairs leave (enter) the superconductor through the cylindrical surface thus decreasing (increasing) in the wire the number of charge carriers participating in the transport along the $z$ axis.

In conclusion of this  section, we note that the complex energy $E$ means, according to (\ref{Temperature1}), that the temperature $T$ becomes complex too. According to our earlier interpretation, real part of the temperature means an actual temperature of the superconductor sample while its imaginary component describes the rate with which the number of the Cooper pairs changes.

\begin{figure}
\centering
\includegraphics[height=7.0in]{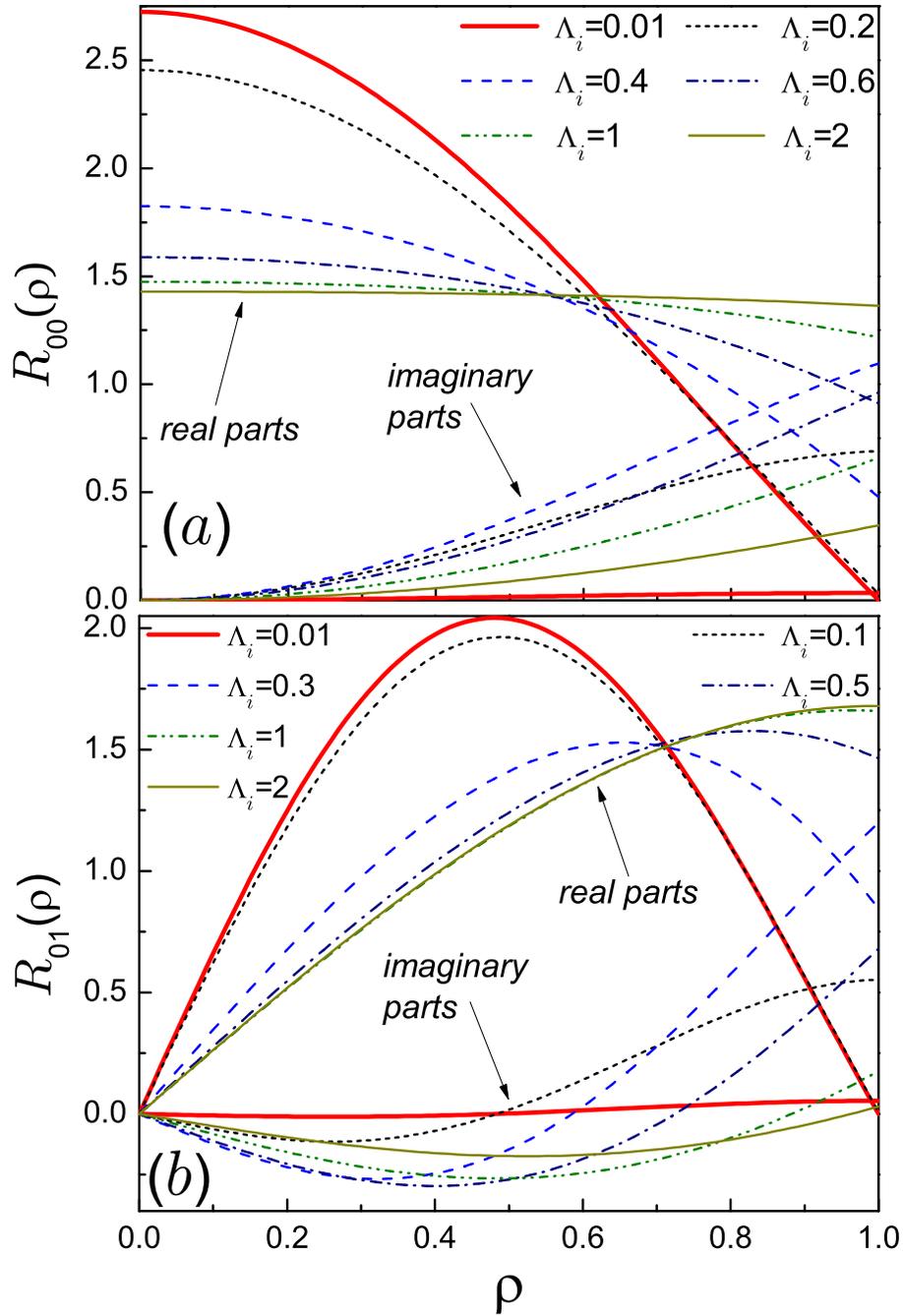}
\caption{\label{Fig4}
Real (upper at $\rho\approx0$) and imaginary (lower at $\rho\approx0$) parts of the total radial functions (a) $R_{00}(\rho)$ and (b) $R_{01}(\rho)$ for several $\Lambda_i$ where for both cases the thick solid lines are for $\Lambda_i=0.01$, dash-dot-dotted lines  - for $\Lambda_i=1$, and thin solid lines are for  $\Lambda_i=2$ while dotted lines are for (a) $\Lambda_i=0.2$ and (b) $\Lambda_i=0.1$,  dashed lines  - for (a) $\Lambda_i=0.4$ and (b) $\Lambda_i=0.3$,  and dash-dotted lines  - for (a) $\Lambda_i=0.6$ and (b) $\Lambda_i=0.5$.
}
\end{figure}

\section{Results and discussion}
\label{sec3}
In this section the results of the calculations based on the theory developed in Section \ref{sec2} are presented and their mathematical and physical interpretations are given.

Fig.~\ref{Fig2} shows $E_{00}^{\left(r\right)}$ and $\Gamma_{00}$ as functions of the pure imaginary de Gennes distance $i\Lambda_i$ and magnetic intensity $B$. It is seen that the real part of the energy as a function of $\Lambda_i$ at the fixed field decreases with the distance growing. As the field increases, the energy dependence on the extrapolation length gets smaller, and at the large field, as was deduced earlier, $E_{nm}^\perp$ tend to the Landau levels, Eq.~(\ref{LandauLevels1}), when the energy is $\Lambda$-independent and real, $\Gamma=0$. However, at the zero and small fields, the imaginary part of the total energy exhibits a pronounced extremum on the $\Lambda_i$-dependence with it absolute value increasing at the decreasing field. Similar dependencies demonstrate also the states with other quantum numbers $n$ and $m$.

Before considering the magnetic field influence, let us discuss first the field-free case. In Fig.~\ref{Fig3} complex energies $E_{nm}^\perp$ are drawn as functions of the imaginary length $i\Lambda_i$ for several $n$ and $m$. Panel (a) shows that for each state $(n,m)$, as $\Lambda_i$ changes from zero, the real part $E_{nm}^{(r)}$ monotonically decreases from $(j_{\left|m\right|n}/\pi)^2$ at the pure Dirichlet case, $\Lambda=0$, to $(j_{\left|m\right|n}'/\pi)^2$ characteristic for the Neumann requirement, $|\Lambda|=\infty$. Here, $j_{\left|m\right|n}$ and $j_{\left|m\right|n}'$ are the $n$th roots of the equations $J_{\left|m\right|}(x)=0$ and $J_{\left|m\right|}'(x)=0$, respectively \cite{Abramowitz1}. The panel inset makes a comparative analysis between the ground-state energy dependence on the positive Robin parameter $\Lambda_r$ and  the real part of the transverse energy dependence on the purely imaginary de Gennes distance $i\Lambda_i$ of the same magnitude. It is seen that in either case the energies monotonically decrease to zero with $|\Lambda|$ growing; however, in the former case the derivative $\partial E_{00}/\partial\Lambda_r$ is a monotonic $\Lambda_r$ function too while for the imaginary distances the derivative $\partial E_{00}^{(r)}/\partial\Lambda_i$ has an extremum on the $\Lambda_i$ axis. Note also the faster approach to zero by $E_{00}^{(r)}$ for the large magnitudes of $|\Lambda|$.

Panel (b) of Fig.~\ref{Fig3} shows that the $\Gamma_{nm}-\Lambda_i$ dependence demonstrates highly nonmonotonic behavior with  characteristic resonance whose sharpness and magnitude increase with $n$ and $|m|$. This prominent feature can be understood as follows. An addition theorem and asymptotic properties of the Bessel functions \cite{Abramowitz1} applied to (\ref{BesselEq1}) lead to the following expressions for $\Gamma_{n\left|m\right|}$ in the limit of the small, $|\Lambda_i|\ll 1$, and large, $|\Lambda_i|\gg 1$, magnitudes of the purely imaginary de Gennes distance:
\begin{equation}\label{AsymptoticsGamma1}
\Gamma_{n\left|m\right|}=\frac{4}{\pi^2}\left\{
\begin{array}{cc}
j_{\left|m\right|n}^2\Lambda_i, &\left|\Lambda_i\right|\ll 1\\
\frac{2J_{\left|m\right|}(j_{\left|m\right|n}')}{\delta_{n0}\delta_{m0}+J_{\left|m\right|+1}'(j_{\left|m\right|n}')-J_{\left|m\right|-1}'(j_{\left|m\right|n}')}\frac{1}{\Lambda_i},&\left|\Lambda_i\right|\gg 1,
\end{array}
\right.
\end{equation}
$\delta_{nn'}$ is a a Kronnecker symbol. Note that the coefficients at both $\Lambda_i$ and $1/\Lambda_i$ are positive; accordingly, as the imaginary part of the extrapolation length moves away from the two extreme points $\Lambda_i=0$ and $1/\Lambda_i=0$, the magnitude of $\Gamma_{nm}$ grows. Matching these two asymptotics in the intermediate regime, $\Lambda_i\lesssim 1$, leads to the pronounced resonance with its maximum increasing with $|m|$ and $n$. Mathematically, this follows from the fact that $j_{|m|n}$ is an increasing function of both $|m|$ and $n$ \cite{Abramowitz1}. This increase is easily explained from the physical point of view too; namely, at the fixed principal number $n$ (azimuthal number $|m|$) the states with the bigger $|m|$ ($n$) are located further from the origin; accordingly, they are influenced stronger by the change in the boundary conditions. In our model, such a resonance means that by a simple change of the applied gate voltage defining the field $\mbox{\boldmath$\cal E$}$ one can control an attenuation of the supercurrent over a wide range, as it follows from (\ref{TotalZCurrent}) and (\ref{Wavevector3}).
\begin{figure}
\centering
\includegraphics[width=0.96\columnwidth]{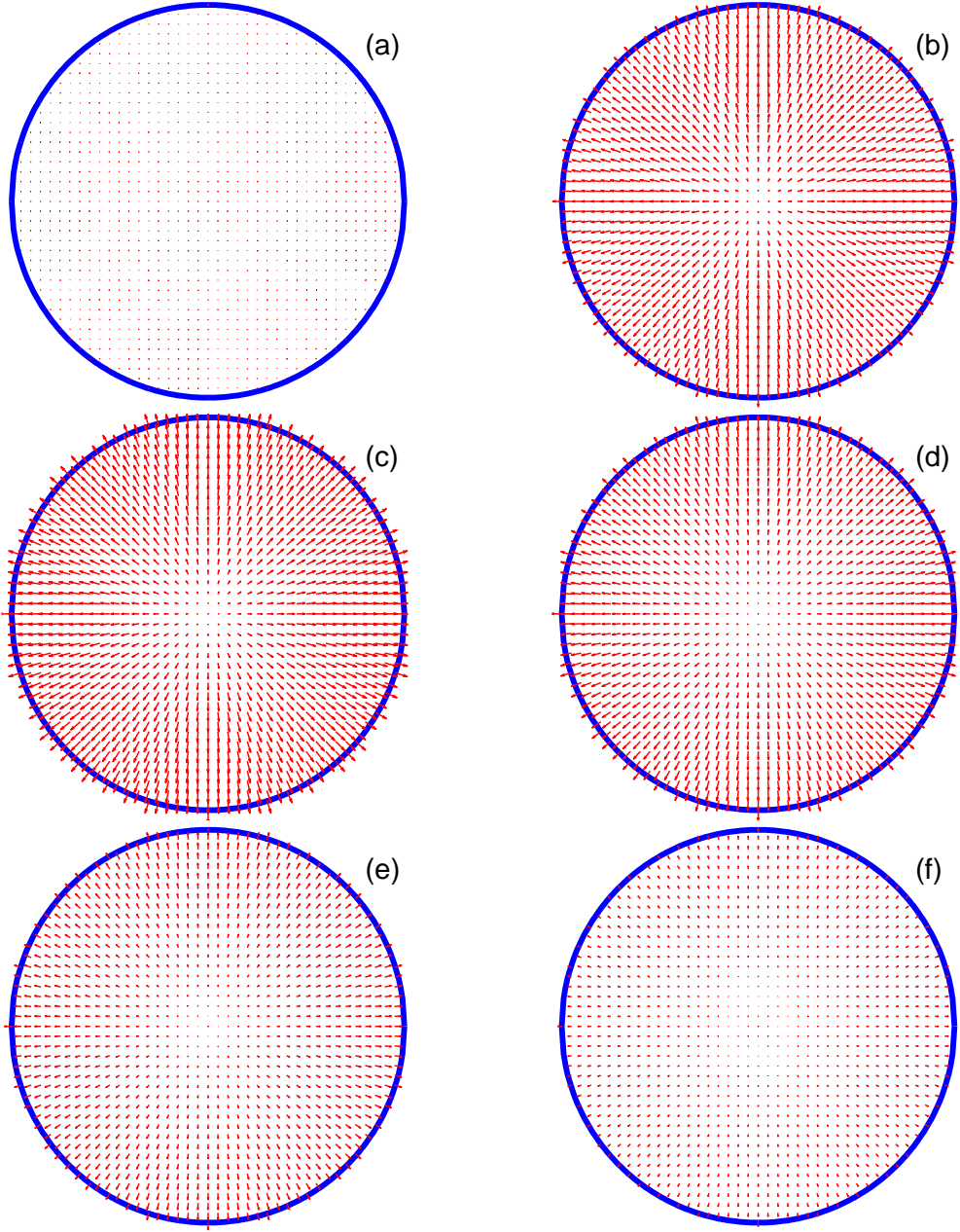}
\caption{\label{Fig5}
Current densities of the state with $n=m=0$ and (a) $\Lambda_i=0.01$, (b) $\Lambda_i=0.2$, (c) $\Lambda_i=0.4$, (d) $\Lambda_i=0.6$, (e) $\Lambda_i=1$, and  (f) $\Lambda_i=2$. Longer arrows denote larger currents.
}
\end{figure}

Fig.~\ref{Fig4} depicts wave functions evolution as the purely imaginary de Gennes distance sweeps its positive axis. Nonzero imaginary part induces complex radial wave function what is a manifestation of the transformation of the true bound state with the infinite lifetime into the quasibound state with finite $\tau_{nm}$. It is seen that the imaginary component of the wave function grows in magnitude for the small $\Lambda_i$, reaches its absolute maximal value approximately at the same de Gennes distance when the maximum of $\Gamma_{nm}$ is achieved and at the further growth of the extrapolation length gradually turns to zero again. At the same time, the real components transform from the Dirichlet dependence at $\Lambda_i=0$ to the Neumann case at the infinite de Gennes distance; in particular, the radial function $R_{00}(\rho)$ approaches a constant value $\sqrt{2}$ for the large $|\Lambda|$.

\begin{figure}
\centering
\includegraphics[width=0.96\columnwidth]{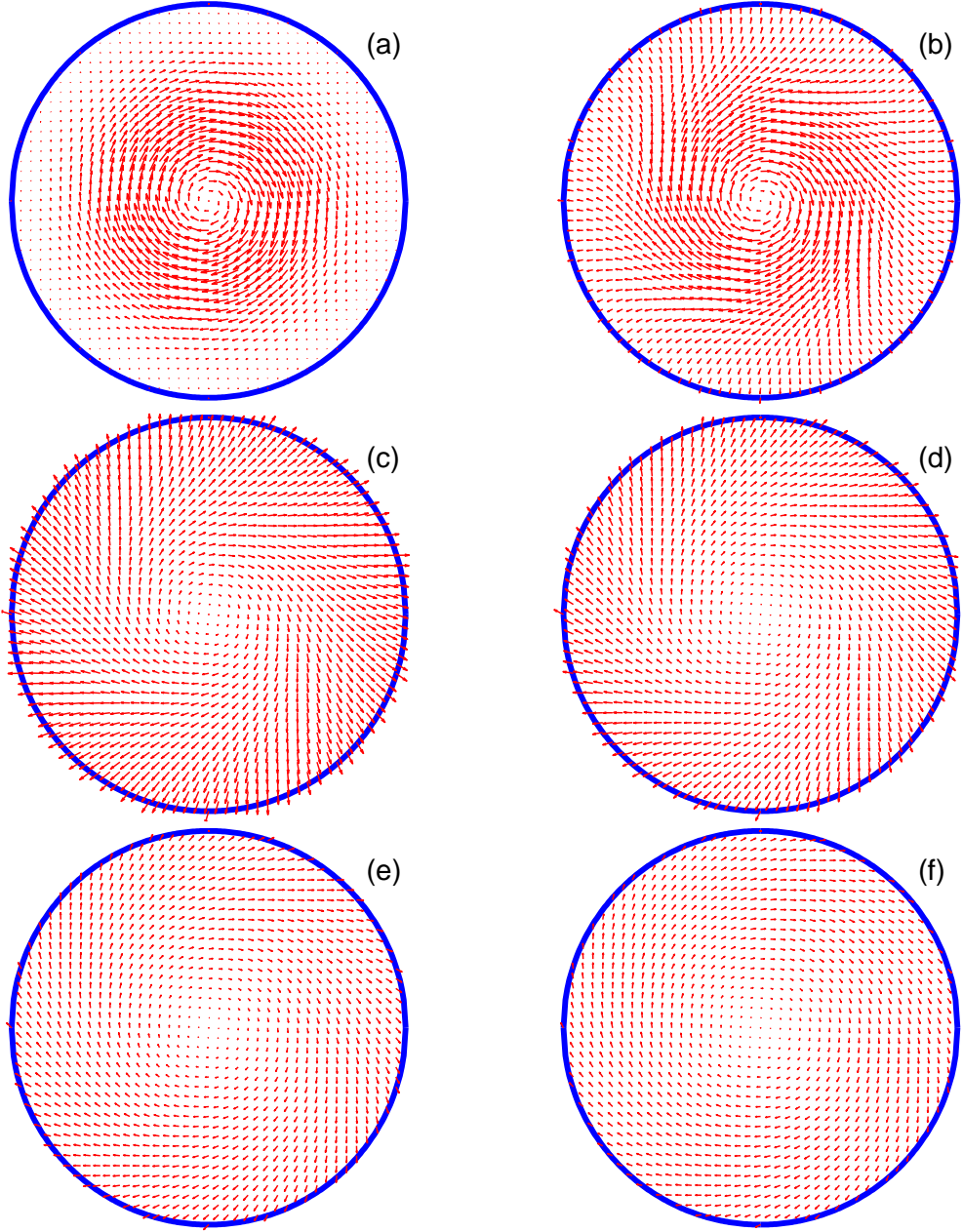}
\caption{\label{Fig6}
The same as in Fig. \ref{Fig5} but for $n=0$, $m=1$ and (a) $\Lambda_i=0.01$, (b) $\Lambda_i=0.1$, (c) $\Lambda_i=0.3$, (d) $\Lambda_i=0.5$, (e) $\Lambda_i=1$, and  (f) $\Lambda_i=2$.
}
\end{figure}

Current density patterns corresponding to the functions from Fig.~\ref{Fig4} are plotted in Figs.~\ref{Fig5} and \ref{Fig6}. They clearly show the emergence and the increase of the normal component of the supercurrent as the imaginary part of the extrapolation length reaches the resonance value and its gradual decrease to zero upon further  increase of $\Lambda_i$. In the absence of the magnetic field, the state with $m=0$ for the two limiting cases $|\Lambda|=0$ and $1/|\Lambda|=0$ does not show any transverse currents at all while for the finite $\Lambda_i$ the radial component emerges, as Fig.~\ref{Fig5} demonstrates. For the nonzero azimuthal numbers the pure rotational whirling with $j_\rho=0$ for the Dirichlet case is distorted at $\Lambda_i\ne0$ by the appearance of the radial motion which pushes carriers out of the sample (Fig.~\ref{Fig6}). Similar to the $m=0$ case, this radial part reaches maximum approximately at the resonance extrapolation length and then with $\Lambda_i$ growing gradually fades to zero transforming the picture into the currents in the Neumann disk. These current density patterns supplement theoretical discussion of the previous section about the carriers flow through the superconductor boundary.

\begin{figure}
\centering
\includegraphics[width=0.75\columnwidth]{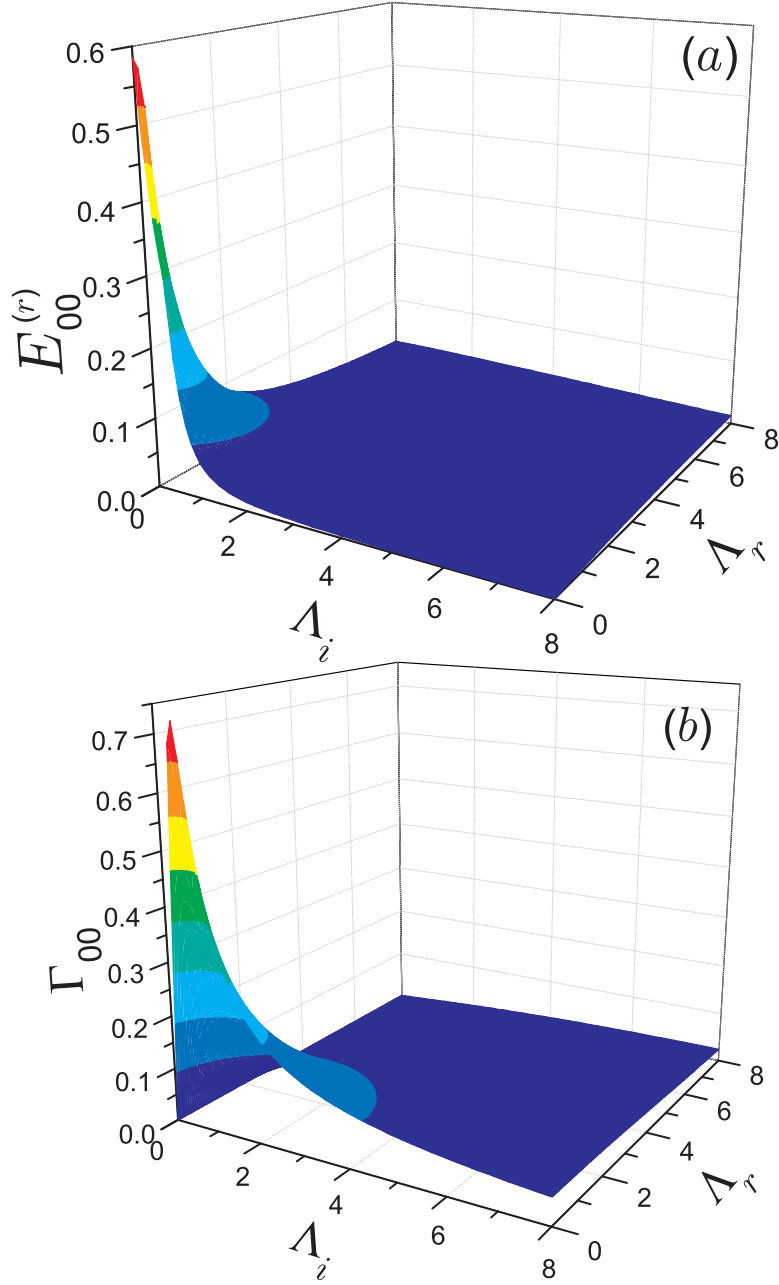}
\caption{\label{Fig7}
(a) Real $E_{00}^{(r)}$ and (b) negative double imaginary $\Gamma_{00}$ parts of the total transverse energy $E_{00}^\perp$ as functions of $\Lambda_i$ and positive $\Lambda_r$ at zero magnetic field.
}
\end{figure}

So far, we considered the case of the zero real part of the de Gennes distance. Fig.~\ref{Fig7} justifies such an approach as its panel (b) shows that the finite positive $\Lambda_r$ strongly quenches the $\Gamma_{nm}$ resonance, and at $\Lambda_r\gtrsim 5$ the imaginary part of the transverse energy almost vanishes and is independent of $\Lambda_i$. For the finite $\Lambda_r$ and small magnitudes of $\Lambda_i$, in the way, similar to (\ref{AsymptoticsGamma1}),  one gets for the halfwidth $\Gamma_{n|m|}$:
\begin{equation}\label{AsymptoticsGamma2}
\Gamma_{n|m|}=\frac{4}{\pi^2}\frac{\lambda_{\left|m\right|n}^2J_{\left|m\right|}'(\lambda_{\left|m\right|n})}{\left(1+\Lambda_r\right)J_{\left|m\right|}'(\lambda_{\left|m\right|n})+\frac{1}{2}\lambda_{\left|m\right|n}\Lambda_r\left[J_{\left|m\right|-1}'(\lambda_{\left|m\right|n})-J_{\left|m\right|+1}'(\lambda_{\left|m\right|n})\right]}\Lambda_i,\qquad\left|\Lambda_i\right|\ll 1,
\end{equation}
where $\lambda_{\left|m\right|n}$ is $n$th root of equation
$$
xJ_{\left|m\right|}'(x)+\frac{1}{\Lambda_r}J_{\left|m\right|}(x)=0.
$$
For the vanishing real parts, $\Lambda_r\rightarrow 0$, the solution $\lambda_{\left|m\right|n}$ tends to $j_{\left|m\right|n}$, and (\ref{AsymptoticsGamma2}) transforms into the corresponding limit of (\ref{AsymptoticsGamma1}). In the opposite regime of $\left|\Lambda_r\right|\rightarrow\infty$ the halfwidth approaches zero with the root $\lambda_{\left|m\right|n}$ changing to $j_{\left|m\right|n}'$. The expression for $\Gamma_{n\left|m\right|}$ in the limit $|\Lambda_i|\gg 1$ depends on the relation between $\Lambda_r$ and $\Lambda_i$ and, in the scale of Fig.~\ref{Fig7}, basically coincides with the lower line of (\ref{AsymptoticsGamma1}). This is seen from Fig.~\ref{Fig7}(b) where for the large $\Lambda_i$ the imaginary part of the energy only slightly depends on $\Lambda_r$. Summing up this discussion, one concludes that the real background of the extrapolation length suppresses the influence of its imaginary component and restores the lossless particle flow without its significant changes in the external electric field. Accordingly, for the best experimental observation of the supercurrent attenuation or amplification, one needs to use ferromagnet as a surrounding material when, as stated in the Introduction, the real part of the de Gennes distance is very close to zero. The effect survives if the superconductor borders the normal metal ($0<\Lambda_r<\infty$) even though its magnitude gets decreased, and completely vanishes for the boundary with the insulator or the vacuum, $1/\Lambda_r=0$, what is quite natural since in these ambient environments the Cooper pairs can not exist at all and, accordingly, they can not enter or leave the superconductor.

As expected, the real part of the energy monotonically decreases to its Neumann counterpart if either part of the complex extrapolation length tends to the large values as is shown in panel (a). For the negative real parts of the Robin parameter the picture of the resonance quenching is basically the same except of the very vicinity of the negative zero when the energy takes unrestrictedly large negative values upon the extrapolation length approach to zero from the left \cite{Berry2,Olendski5}. In this case the increasing imaginary part promotes the negative energy $E_{00}^{(r)}$ upwards to its zero Neumann case; however, at the very small negative $\Lambda_r$, that energy could be occupied by the level with $n=1$. Repulsive interaction between these two states leads to the anticrossing between  $E_{00}^{(r)}$  and $E_{10}^{(r)}$ on the $\Lambda_i$ axis with its sharpness dependent on the smallness of the negative $\Lambda_r$ and results in the repel of the higher lying state from the lowest Neumann position. To make our exposition more compact, we do not present this transformation here.

It is instructive to compare Figs.~\ref{Fig2} and \ref{Fig7}. Panels (b) show quite a similarity, at least qualitative, while the real parts of the transverse energy exhibit completely different behavior. Thus, in order to observe the most profound supercurrent alteration, one needs to use materials with $\Lambda=0$ and, to check the quenching of the resonance,  it is necessary to turn the external magnetic field on.

\begin{figure}
\centering
\includegraphics[width=0.75\columnwidth]{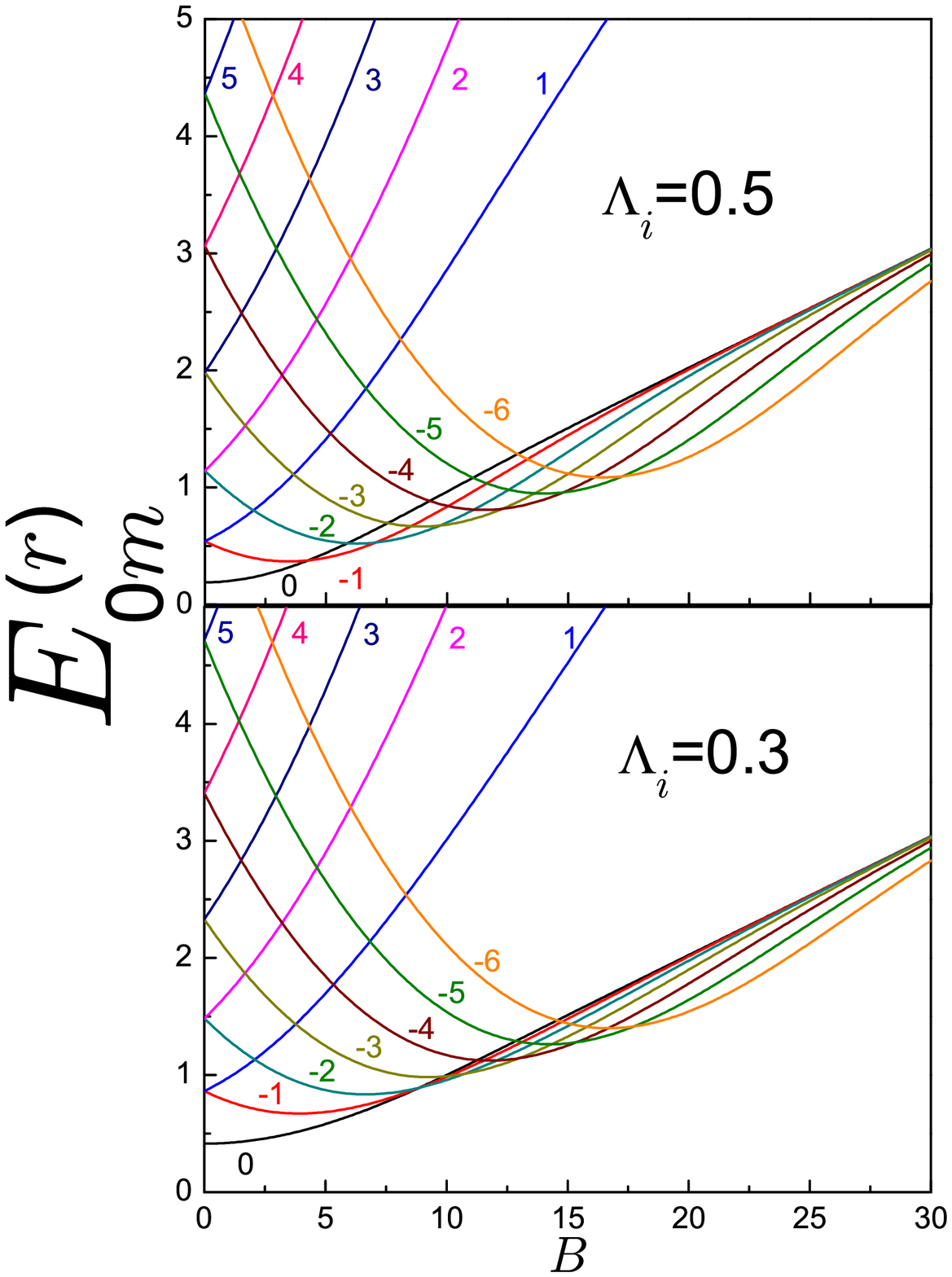}
\caption{\label{Fig8}
Real parts $E_{0m}^{(r)}$ of the total transverse energies $E_{0m}^\perp$ as functions of the magnetic field $B$ for $\Lambda_i=0.5$ (upper panel) and $\Lambda_i=0.3$ (lower panel). Numbers near the curves denote corresponding azimuthal quantum numbers $m$.
}
\end{figure}

The last paragraph naturally returns us to the study of the combined influence of the magnetic field and complex extrapolation length. We have already seen in Fig.~\ref{Fig2}(b) that the increasing magnetic field washes out the resonance on the $\Gamma-\Lambda_i$ axis since it pushes the Cooper pairs closer to the wire axis and thus restricts their interaction with the disturbing borders of the wire. For the very large magnetic fields all the states $(n,m)$ approach from below the corresponding Landau levels, Eq.~(\ref{LandauLevels1}), what means their independence on the imaginary component $\Lambda_i$. It is known that for the real extrapolation length similar asymptotics of the Landau levels crucially depends on $\Lambda_r$; namely, it was calculated that for the Dirichlet confinement the state with $m=0$ at any magnetic intensity remains the lowest one  of the whole family of the levels with the same principal quantum number $n$ \cite{Nakamura1,Geerinckx1}. Situation is very different for the Neumann case, $1/\Lambda_r=0$, where the levels with the adjacent negative azimuthal quantum numbers $-|m|$ and $-|m|-1$ cross with the field increasing \cite{Buisson1,SaintJames1,Moshchalkov2,Benoist1} thus causing the ground state to change its vorticity. These theoretically predicted crossings were indeed observed experimentally for the mesoscopic aluminium  disks \cite{Moshchalkov1,Geim1,Chibotaru1}. The crossings are characteristic for any nonzero extrapolation length with their location shifted to the higher magnetic fields for the smaller de Gennes distance until at $\Lambda=0$ they disappear and the state $(0,0)$ remains the ground state for any $B$ \cite{Buisson1}. Fig.~\ref{Fig8} showing the dependence of the real part of the transverse energy on the magnetic field  for the purely imaginary distances $\Lambda_i=0.3$ and $\Lambda_i=0.5$ confirms this property for the complex extrapolation lengths too. These changes by the ground state of its azimuthal number are responsible for the jumps of the magnetization.

\begin{figure}
\centering
\includegraphics[width=0.75\columnwidth]{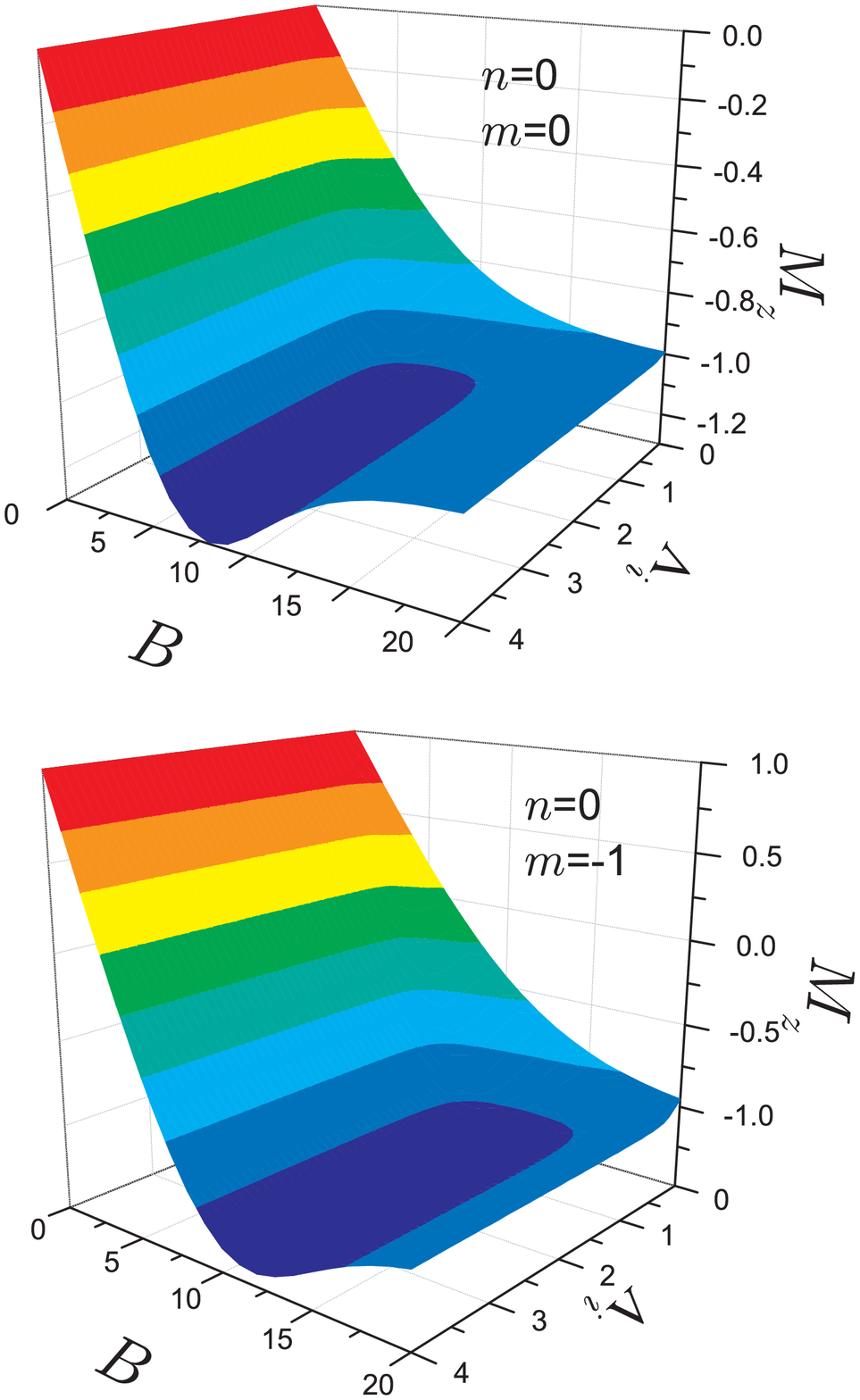}
\caption{\label{Fig9}
Magnetizations $M_z$ of the states $(0,0)$ (upper panel) and $(0,-1)$ (lower panel) as functions of magnetic field $B$ and purely imaginary de Gennes distance $i\Lambda_i$. Note different $M_z$ scales for the two panels.
}
\end{figure}

Magnetizations $M_z$ for the states $(0,0)$ and $(0,-1)$ are shown in Fig.~\ref{Fig9} as functions of the magnetic field and imaginary de Gennes distance $i\Lambda_i$. The magnetic moments from its zero-field value $-m$ [see Eq.~(\ref{Magnetization3}) for $B=0$], decrease for the large $B$  to the magnetization of the free charged particle in the uniform magnetic field $M_z=-(2n+|m|+m+1)$. For the Dirichlet disk this approach is a monotonic function of the intensity while for the Neumann case the magnetization reaches minimum as a function of the field after which it monotonically increases to the saturation value. Extremum location on the $B$ axis increases with the azimuthal number $|m|$. A transformation between the two cases with $\Lambda_i$ changing is clearly seen in the figure. Contrary to the longitudinal supercurrent features, the dependencies in Figs.~\ref{Fig8} and \ref{Fig9} are very similar to the case of the real de Gennes lengths. Accordingly, magnetization measurements are hardly suitable for the detection of the complex boundary conditions.


\section{Concluding remarks}
\label{sec4}
The analysis of the model of the superconducting 2D disk or infinitely long 3D cylinder with the nonzero imaginary part of the Robin parameter confirmed the general quantum mechanical rule that the complex boundary condition leads to the complex eigenenergies of the corresponding Shcr\"{o}dinger equation and induces currents through the superconductor interface into the surrounding medium. For the case of the infinite wire these transverse leaking currents alter the longitudinal flow of the Cooper pairs with the resonance dependence on the imaginary part of the de Gennes distance.

As was stated in the Introduction, the present treatment for the zero magnetic field can be directly applied to the study of the dynamics in the acoustical, impedance and ferrite-filled electromagnetic waveguides as well as the processes in the superconducting structures close to the transition between the normal and superconducting states. The superconductors in the magnetic field are also described by the theory developed above. As was mentioned in Sec.~\ref{sec2}, for the thin superconducting disks the results obtained are supported by the nonlinear GL theory as well. Moreover, numerical simulation \cite{Deo1} of the aluminum disks with the Neumann requirement revealed that the predictions of the linearized GL equation are in a better agreement with the experiment \cite{Geim1} than the full GL treatment. Furthermore, theoretical linear GL analysis \cite{Deo1} correctly captures experimental findings \cite{Geim1} down  to temperatures $T=0.4$ K lying well below corresponding $T_c=1.19$ K for aluminum. Accordingly, there is a strong confidence that the resonances discovered in the present investigation, could be experimentally detected in the superconducting guiding structures with the applied external electric field tuning the total complex extrapolation length $\Lambda$.

The de Gennes boundary condition uses only {\it one} parameter $\Lambda$ for the description of the interaction between the superconductor and surrounding medium and ignores the processes occurring outside the superconducting sample. However, the Cooper pairs can penetrate into the adjacent normal metal and, obviously, into the neighboring superconductor with the higher critical temperature. This can be especially important in our case of the complex boundary conditions when these charge carriers flowing outside the actual superconductor can contribute to the total longitudinal supercurrent and magnetization. Modified GL theory taking into account nonzero order parameter in the bordering normal metal was recently proposed \cite{Chapman1} and developed \cite{Giorgi4,Kachmar6,Kachmar7,Fournais1}. Instead of the {\it single} parameter $\Lambda$,  it utilizes the {\it two} ones: the carrier mass $\tilde{m}$ (here we operate in regular, dimensional units and use tilde for the values of the modified GL theory) which is different in the normal metal with its value of $\tilde{m}_n$ and in the superconductor ($\tilde{m}_s$), and the dimensionless parameter $a=\left|(\tilde{m}_n\tilde{\alpha}_n)/(\tilde{m}_s\tilde{\alpha}_s)\right|$ where the GL parameter $-\tilde{\alpha}$ has different signs inside and outside the superconductor, $\tilde{\alpha}_s\tilde{\alpha}_n<0$. As a result, it requires on all the interfaces a continuity of the vector potential $\tilde{\bf A}$, order parameter $\tilde{\Psi}$, and values of ${\bf n}\frac{1}{\tilde{m}}\left(i\hbar{\bm\nabla}-q{\tilde{\bf A}}\right)\tilde{\Psi}$ and $\frac{1}{\tilde{\mu}}\left({\bm\nabla}\times\tilde{{\bf A}}\right)\times{\bf n}$ with $\tilde{\mu}$ being the global magnetic permeability. Taking into account the currents flowing outside the superconductor can alter the results presented here. However, this modification should not smear out the resonances at all, especially for the case of the bordering ferromagnets which effectively destroy superconductivity by breaking the Cooper pairs. In a sense, the model of the complex extrapolation length is similar to the modified GL theory since both of them use the two parameters for the description of the superconductor/other medium interface (in our case these are the real and imaginary components of the de Gennes distance) and allow the current flow across the border.

For considered here system of the quantum disk there is only one surface through which the energy can flow into  or out from the structure. In turn, for the quantum ring with the interfaces of the smaller and larger radii there are two such channels. Intuitively, one can expect that for some correlation between the inner and outer complex de Gennes distances with the opposite signs of their imaginary parts the energy can become real again. Apparently, this happens when the flux entering through one of the surfaces is equal in magnitude to the outgoing flow through the other boundary. For the rectangular geometry it is elementary to show that it happens when the de Gennes lengths on the opposite walls are complex conjugate (see Eq.~(10) in Ref. \cite{Olendski5}). However, for the annulus due to the centrifugal forces \cite{Nesvizhevsky1} this relation does  not seem to be so trivial. This case requires separate special consideration.

\bibliographystyle{model1a-num-names}

\end{document}